\documentclass[5p]{elsarticle}
\usepackage{amsmath,amssymb,color}
\usepackage{graphicx, graphics, hyperref}
\usepackage{psfrag}
\usepackage{textcomp}

\newcommand{\prl}{{Phys.~Rev.~Lett.~}}
\newcommand{\prd}{{Phys.~Rev.~D} }
\newcommand{\pre}{{Phys.~Rev.~E} }

\begin{document}

\title{Relative information entropy in cosmology: The problem of information entanglement}

\author[cmat,wigner]{Viktor G.~Czinner}
\ead{czinner.viktor@wigner.mta.hu}
\author[cmat]{Filipe C. Mena}
\ead{fmena@math.uminho.pt}
\address[cmat]{Centro de Matem\'atica, Universidade do Minho, Campus de Gualtar, 
4710-057 Braga, Portugal}
\address[wigner]{HAS Wigner Research Centre for Physics, H-1525 Budapest, P.O.~Box 49, Hungary}

\date{\today}

\begin{abstract}
The necessary information to distinguish a local inhomogeneous mass density field from its spatial 
average on a compact domain of the universe can be measured by relative information entropy.
The Kullback-Leibler (KL) formula arises very naturally in this context, however, it provides
a very complicated way to compute the mutual information between spatially separated but causally 
connected regions of the universe in a realistic, inhomogeneous model. To circumvent 
this issue, by considering a parametric extension of the KL measure, we develop a simple model to describe 
the mutual information which is entangled via the gravitational field equations. We show that 
the Tsallis relative entropy can be a good approximation in the case of small inhomogeneities, and 
for measuring the independent relative information inside the domain, we propose the R\'enyi relative
entropy formula.
\end{abstract}



\maketitle

\section{Introduction}
Understanding the evolution of local and global inhomogeneities of the universe is a central 
problem in modern cosmology. The question has been actively investigated from several different 
points of view (e.g.~perturbation theory \cite{pert}, exact models \cite{exact}, numerical 
simulations \cite{simul}), and a more recent approach attacks from the direction of information 
theory \cite{HBM,AC}. In the work of Hosoya {\it et al}.~it has been shown \cite{HBM}, that 
the Kullback-Leibler relative information entropy \cite{KL} -- a standard notion in information 
theory -- arises very naturally in this context. It can describe the "distinguishability" of an 
actual mass density field from its spatial average on a compact domain of the universe, and its 
time derivative can also provide the commutation relation between the time evolution and the volume 
average operation on the density field. 

The KL entropy is therefore a very useful measure to study the evolution of cosmic inhomogeneities
inside a given domain, however, since gravity is a long-range interaction, it is often the case 
that the mutual information between spatially separated but causally connected domains of the 
universe is also important to consider. In order to compute the entangled information via the 
gravitational field equations in a realistic, inhomogeneous model, one has to face a rather 
formidable problem. The exact way to do so would be to follow a dynamical volume partitioning 
method (see e.g.~the work of Wiegand and Buchert \cite{WB}), where the matter distribution
function would have to be known globally, by solving complicated system of the Einstein's 
field equations. From an observationally motivated point of view for example, this 
approach is particularly difficult.

To circumvent this issue, in this Letter, we propose a simple model where
we consider a parametric extension of the KL measure to describe the evolution of cosmic 
inhomogeneities on a domain whose dynamics is informationally entangled to its causally connected 
surroundings. In order to extract the mutual information between separated domains, instead of 
computing the explicit relations between the distribution functions via the exact method above,
we consider a different approach where the distribution functions are formally treated as independent, 
but the corresponding relative entropy functions are nonadditive. We show, that under some physically 
motivated assumptions,
this approach leads quite generally to the Tsallis relative entropy formula \cite{Tsallis2}, which 
can parametrically incorporate the long-range interaction property of the gravitational field, based 
on the causal structure of the problem. Although nonadditive (nonextensive) phenomena have been known 
in cosmology and gravitation theory for decades, and the standard Tsallis entropy \cite{Tsallis1} has 
also been investigated several times in cosmological applications \cite{TSL}, as far as we know, this 
is the first time that a similar approach is considered in connection with cosmic inhomogeneities and 
information entropy.  For simplicity, in the present work we restrict our investigations to linearly
perturbed, spatially flat, Friedmann-Lema\^{\i}tre-Robertson-Walker (FLRW) dust cosmologies, however we 
expect that our model can be extended to more general, inhomogeneous cosmological models as well. 
Throughout this Letter we use units such that $c=G=1$.

\section{Background} 
The KL relative information entropy for continuous 
probability distributions $p(x)$ and $\overline{p}(x)$ on a compact domain, $D$,
is defined as 
\begin{equation}
\label{SKL}
 S_{KL}\{p\!\mid\!\overline{p}\}=\int_D p(x)\ln\frac{p(x)}{\overline{p}(x)}dx.
\end{equation}
For studies of cosmic inhomogeneities, the relevant distributions to consider 
are the matter density $\rho(t,x^i)$ and its spatial average, $\overline{\rho}(t)$,
over an arbitrary domain of the universe. Hosoya {\it et al}.~\cite{HBM} have shown, 
that for a dust continuum cosmological model, described by the metric
\begin{equation}
ds^2=-dt^2+g_{ik}dx^idx^k,
\end{equation}
with a predefined time-orthogonal foliation, 
the KL relative entropy can be written as 
\begin{equation}
\frac{S_{KL}\{\varrho\!\mid\!\overline{\varrho}\}}{V_{D}} 
\ =\ \overline{\varrho\ln\frac{\varrho}{\overline{\varrho}}}\ ,
\end{equation}
where overbar denotes the volume average operation: 
\begin{equation}\label{ave}
 \overline{\psi}(t)= \frac{1}{V_{D}}\int_{D}\psi(t,x^i)\sqrt{g}\,d^3x,
\end{equation}
defined for any scalar field $\psi$ in the volume $V_{D}=\int_{D}\sqrt{g}d^3x$.
In (\ref{ave}) $g$ is the determinant of the 3-metric $g_{ik}$, and $x^i$ are 
coordinates on a $t=\mbox{const.}$ hypersurface within a comoving gauge.
By exploiting the identity 
$\dot{\overline{\psi}}-\overline{\dot\psi}\,
=\,\overline{\psi\,\theta}-\overline{\psi}\,\overline{\theta}$
with $\theta=\dot V_{D}/V_{D} $ being the local expansion rate,
and also applying the continuity equation
$ \dot{\overline{\varrho}}\,+\,\overline{\theta}\,\overline{\varrho}
\,=\,\overline{\dot\varrho\,+\,\theta\varrho}\,=\,0,$
it can be shown \cite{HBM} that the KL relative entropy is the generating functional of 
the commutation relation between the volume average and the time evolution of 
the density field, in the sense that
\begin{equation}\label{dSKL}
 -\frac{\dot S_{KL}\{\varrho\!\mid\!\overline{\varrho}\}}{V_{D}}
=\dot{\overline{\varrho}}-\overline{\dot\varrho}\ .
\end{equation}
Hosoya {\it et al}.~analyzed this relation, and argued 
that after long enough time, and on sufficiently large scales of averaging, 
$S_{KL}\{\varrho\!\mid\!\overline{\varrho}\}$ is an increasing function of
time, and hence a reasonable entropy description for the evolution of local
inhomogeneities.

\section{The causal structure of the problem} 
\begin{figure}[!ht]
\includegraphics[
scale=.43]{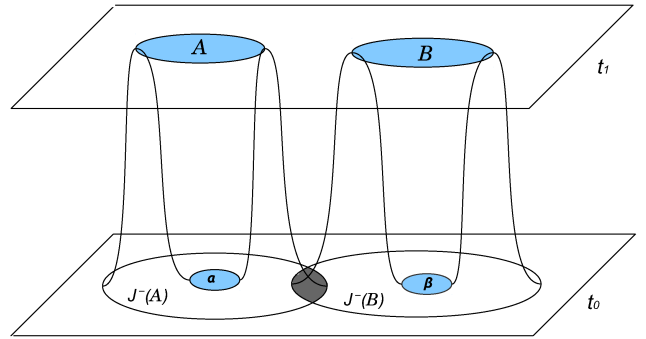}
\caption{\label{Fig1}
The causal structure of the evolution of two spatially separated domains in an FLRW universe. 
$\alpha$ and $\beta$ are the cosmological pasts of $\mathcal{A}$ and $\mathcal{B}$ respectively, 
while $J^{-}(\mathcal{A})$ and $J^{-}(\mathcal{B})$ are their intersecting causal pasts on $t_0$.}
\end{figure}

The KL relative information entropy 
is additive for factorizing probabilities. More precisely, 
$S_{KL}\{\mathcal{A}\otimes\mathcal{B}\}=S_{KL}\{\mathcal{A}\}+S_{KL}\{\mathcal{B}\}$,
if $\mathcal{A}$ and $\mathcal{B}$ are two {\it independent} systems in the sense that the
probability distributions $p(x)$ and $\overline{p}(x)$ of $\mathcal{A}\otimes\mathcal{B}$ 
{\it factorizes} into those of $\mathcal{A}$ and of $\mathcal{B}$. As a consequence, the 
{\it mutual information}, defined usually as
\begin{equation}\label{mi}
I(\mathcal{A},\mathcal{B})\ =\ S\{\mathcal{A}\}+S\{\mathcal{B}\}-S\{\mathcal{A}\otimes\mathcal{B}\},
\end{equation}
is zero for composite independent systems. The assumption of independence, however, is obviously not 
valid for gravitationally coupled systems, like those in cosmology, where the evolution of a density
field on an arbitrary compact domain of the universe is always influenced by its dynamics on its 
neighboring, causally connected regions via the gravitational field equations. 
On Fig.~\ref{Fig1}, we schematically plotted the causal structure of the evolution for an FLRW model. 
Here $J^{-}(\mathcal{A})$ is the causal past of $\mathcal{A}$ on an arbitrary reference time slice 
$t_0$ in the past, while $\alpha$ is the cosmological past of $\mathcal{A}$. $\mathcal{B}$ is a 
spatially separated domain from $\mathcal{A}$ on $t_1$, $\beta$ is its cosmological-, while
$J^{-}(\mathcal{B})$ is its causal past on $t_0$. It is clear from the picture that due to the
long-range interaction property of the gravitational field (manifested by the intersection of 
$J^{-}(\mathcal{A})$ and $J^{-}(\mathcal{B})$), the mutual information $I(\mathcal{A},\mathcal{B})$ 
at $t_1$ is obviously not zero,  however as we have pointed out in the introduction, it is complicated
to compute its value in general via the gravitational field equations in a realistic, inhomogeneous 
universe model by using an exact dynamical volume partitioning method, like the one in \cite{WB}. 
Nevertheless, since the time evolution of a density field on a domain $D$ 
is clearly not independent from its evolution on causally connected regions to $D$, it would be nice 
to have a simple way to estimate the information entanglement between them. 

In the following sections, by considering a universe model described by a linearly perturbed FLRW 
metric with a dust continuum matter filed, we develop a parametric approximation for the entanglement 
problem. In this model we will stay completely inside the framework defined by Hosoya et al., 
i.e.~we have a global, time orthogonal foliation with an inhomogeneous spacelike metric $g_{ik}$ 
that is comoving with the matter perturbations in the considered linear order.

\section{A parametric entropy extension} 
On large enough scales, as nicely confirmed by CMBR experiments \cite{CMBR}, the universe 
can be well described in a thermodynamic equilibrium during its evolution. Unlike in standard 
thermodynamics however, as we have seen in the previous section, spatially separated domains 
do not evolve independently in general relativity, so it is a natural consequence 
that the entropy function of these regions are {\it not additive} for composition. When computing 
the joint entropy of separated domains in the model of Hosoya et al., the nonadditive 
part in the KL measure arises from the mutual dependence of the density and metric functions of 
the domains via the gravitational interaction. The practical computability of this entanglement 
is very difficult, and our approach in this Letter is to develop a toy model instead, where we 
formally treat the density functions as independent, but consider a parametric extension of the 
KL entropy function which is nonadditive even for independent distributions. The new entropy parameter 
will then be used to describe the mutual information between causally connected regions of the
universe.

Based on the concept of composability alone, 
Abe show\-ed \cite{Abe}, that the most general nonadditive entropy composition rule which 
is compatible with equilibrium requirements can be written in the form
\begin{equation}\label{Abe}
H_{\lambda}(S_{\mathcal{A}\otimes\mathcal{B}})=H_{\lambda}(S_{\mathcal{A}})+H_{\lambda}(S_{\mathcal{B}})
+{\lambda} H_{\lambda}(S_{\mathcal{A}})H_{\lambda}(S_{\mathcal{B}}), 
\end{equation}
where $H_{\lambda}$ is some differentiable function of $S$, ${\lambda}\in\mathbb{R}$ is a parameter,
and $S_{\mathcal{A}}$, $S_{\mathcal{B}}$ and $S_{\mathcal{A}\otimes\mathcal{B}}$ are 
the entropies of the subsystems and the joint system, respectively. 

For the gravitationally coupled problem of cosmological inhomogeneities, the explicit
form of $H_{\lambda}$ is virtually impossible to compute, not to mention the entropy 
problem of the gravitational field itself. On the other hand, the most recent results 
of large scale structure measurements indicate that the matter distribution in the 
physical (approximately FRLW) universe can be extremely well approximated 
by a homogeneous and isotropic dust model above the $\sim 115Mpc$ scale \cite{Scrim}. 
Local inhomogeneities start to grow below this scale and their distribution becomes 
increasingly structured on smaller and smaller scales. Above this scale, no further 
contribution to the relative information entropy can be expected from the statistically 
homogeneous matter distribution, and for scales not much smaller than this (i.e.~around 
the linear regime in inhomogeneities) the entropy composition rule is not expected 
to deviate too much from additivity. These considerations imply that the ${\lambda}$-parameter 
in Abe's formula is not zero but reasonably small, i.e.~$|{\lambda}|<<1$, 
and also that the deviation of $H_{\lambda}$ from the identity function may not be too 
strong either. For regions with this magnitude of inhomogeneities, the nonadditive 
entropy composition rule in (\ref{Abe}), in the leading order of the ${\lambda}$-parameter, 
can be approximated by the formula
\begin{equation}\label{stnadd}
S_{\mathcal{A}\otimes\mathcal{B}}=S_{\mathcal{A}}+S_{\mathcal{B}}
+{\lambda}S_{\mathcal{A}}S_{\mathcal{B}}+\dots\ .
\end{equation}
This approximate form is not immediate to derive from (\ref{Abe}) under our conditions, 
and we intend to present a detailed proof in a forthcoming communication \cite{CzM}. 

Formula (\ref{stnadd}), on the other hand, is a well known expression, called the Tsallis--Pareto 
composition rule, and the corresponding relative entropy measure has been developed by
Tsallis in \cite{Tsallis2}. The general definition is given by 
\begin{equation}\label{st}
S_T\{p\!\mid\!\overline{p}\}
=\frac{1}{q-1}\int_D p(x)\left[\left(\frac{p(x)}{\overline{p}(x)}\right)^{q-1}\!\!\!-1\right]dx,
\end{equation}
where $q=1+{\lambda}$, and the $q\rightarrow 1$ (${\lambda}\rightarrow 0$) limit recovers the KL 
relative entropy,
as is expected by consistency requirements. Therefore, the Tsallis extension to the KL relative 
entropy formula arises very naturally from Abe's formula (\ref{Abe}) as a consequence of the 
cosmological setup for small inhomogeneities in an FLRW universe, and it is easy to show that 
in our framework it takes the form 
\begin{equation}\label{ST}
\frac{S_T\{\varrho\!\mid\!\overline{\varrho}\}}{V_{D}}
\ =\ \frac{1}{V_{D}}\int_{D}\varrho\,\delta_{\lambda} \sqrt{g}\,d^3x
\ \equiv\ \overline{\varrho\,\delta_{\lambda}},
\end{equation}
where we defined the \emph{${\lambda}$-deformed density contrast} function as
\begin{equation}
\delta_{\lambda}=\frac{1}{{\lambda}}\left[\left(\frac{\varrho}{\overline{\varrho}}\right)^{\lambda}-1\right],
\qquad \lim_{{\lambda}\rightarrow 0}\delta_{\lambda}=\ln(1+\delta),
\end{equation}
with $\delta=(\varrho-\overline{\varrho})/\overline{\varrho}$ being the standard density contrast.

\section {A geometric model for $\lambda$} 
The $q$- or ${\lambda}$-parameter in the Tsallis formula is usually 
constant in different physical situations and its explicit value is a part of the problem 
to be solved. In our model however, we will consider a novel approach by allowing it
to be time dependent in order to describe all possible causal relations between spatially 
separated domains during their evolution. 
In Abe's work \cite{Abe}, the ${\lambda}$-parameter 
appears originally as a separation constant in integrating the problem of the joint entropy
function, hence, it seems to be a natural choice to connect it to some common 
but still independent property of the cosmological domains, i.e.~their common causal past. 
Furthermore, we also expect that the mutual information between the causally connected 
domains should be proportional to the volume of their past causal intersection, so 
let us define
\begin{equation}\label{lambda(t)}
 {\lambda}(t)\ :=
 \ -{\lambda}_0\frac{V_{J^{-}(\mathcal{A})\cap J^{-}(\mathcal{B})}}{V_{J^{-}(\mathcal{A})}
 +V_{J^{-}(\mathcal{B})}-V_{J^{-}(\mathcal{A})\cap J^{-}(\mathcal{B})}}, 
\end{equation}
where $V_{J^{-}(\mathcal{A})\cap J^{-}(\mathcal{B})}$ is the volume of the intersection 
of the causal pasts of $\mathcal{A}$ and $\mathcal{B}$ on $t_0$ (see Fig.~\ref{Fig1}). 
For our interest, $t_0$ can be e.g.~the time of decoupling, ever since the dust 
model is considered to be a reasonable cosmological approximation.

\begin{figure}[!ht]
\begin{center}
\includegraphics[scale=.65]{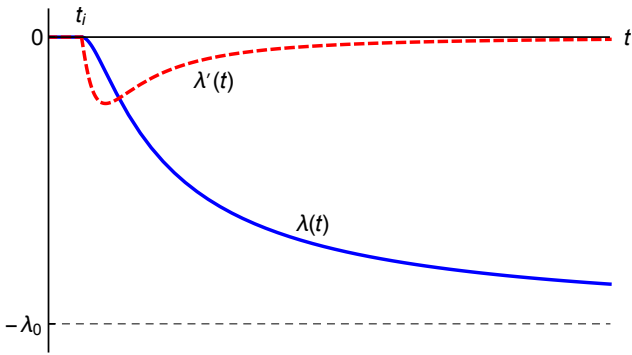}
\end{center}
\caption{ \label{Fig2}
The time dependence of the ${\lambda}(t)$-parameter (blue, continuous curve) and its time derivative 
(red, dashed curve) for the case of two, initially spherical domains $\mathcal{A}$ and $\mathcal{B}$. 
$t_i\in[t_0,t_1]$ denotes the moment when the causal pasts $J^{-}(\mathcal{A})$ and 
$J^{-}(\mathcal{B})$ starts to intersect on $t_0$. The effects of inhomogeneities in the volume ratio 
are neglected on the plot.}
\end{figure}

It is easy to see from the definition that $\lambda(t)$ is zero for causally disconnected 
regions and $\lim_{t\rightarrow\infty}\lambda(t)=-{\lambda}_0$, so ${\lambda}(t)\in (-{\lambda}_0,0]$ 
for all $t$. The ${\lambda}_0$ constant is expected to be different for different regions, and 
its explicit value may be constrained by observational data. On the other hand, based 
on our linear approximation in the model, we require ${\lambda}_0\ll 1$ just below the 
$115Mpc$ scale. On gradually decreasing scales, ${\lambda}_0$ is expected to grow until it reaches 
the limit where our perturbative approach will eventually fail to be satisfied.
Between the two extrema of ${\lambda}(t)$, there is an initial period for most spatially separated 
domains from $\mathcal{A}$, during which their causal past is disjoint from $J^{-}(\mathcal{A})$. 
After this, the explicit form of ${\lambda}(t)$ depends on the
actual geometry of the domains in question. As an example, on Fig.~\ref{Fig2}, we have 
plotted the time dependence of ${\lambda}(t)$ (and its time derivative) for the case of two,
initially spherical domains.  

The interpretation of ${\lambda}(t)$ is fairly straightforward, although it is interesting
to note that within this model, the meaning of relative information entropy 
becomes a {\it relativistic} notion. In other words, it makes sense 
to talk about the relative information entropy of a domain only when we also specify a
{\it reference domain} with respect to which we measure it.  
The ${\lambda}(t)$-parameter keeps track of the causal connection between the domains, and 
via the Tsallis formula we can also compute the mutual information between them.
By substituting (\ref{stnadd}) into (\ref{mi}) we get
$I_T(\mathcal{A},\mathcal{B})=-{\lambda}S_T\{\mathcal{A}\}S_T\{\mathcal{B}\}$, 
which is always positive for causally connected regions, and zero otherwise. 
It can also be checked that the Tsallis relative entropy is a well defined measure 
for all ${\lambda}(t)$, i.e.~it satisfies the necessary conditions: 
$S_T\{\varrho\!\mid\!\overline{\varrho}\}\geq 0$ for $\varrho>0$, and 
$S_T\{\varrho\!\mid\!\overline{\varrho}\}=0$ for $\varrho=\overline{\varrho}$, 
i.e.~for homogeneous distributions the relative information entropy is zero 
\cite{Tsallis2}.

\begin{figure}[!ht]
\includegraphics[
scale=.43]{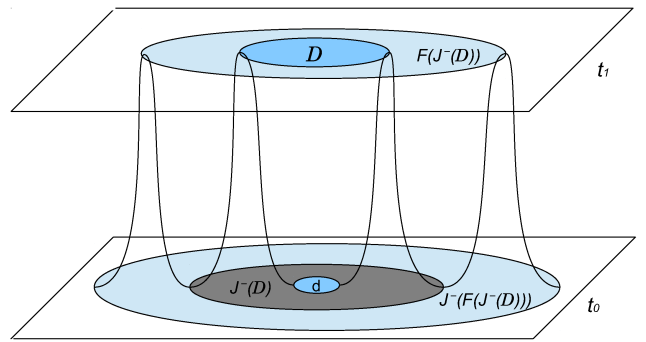}
\caption{\label{Fig3}
The causal structure of the evolution of a cosmological domain and its surroundings.}
\end{figure}

Unless one is specifically interested in estimating the mutual information between, 
say, two disjoint superclusters, the most natural choice for a reference region to 
any domain is its entire causally connected surroundings. According to a result of 
Stewart and Walker \cite{Stewart}, the relative information 
entropy functionals in our linear approximation are gauge invariant since their values 
are identically zero on the homogeneous background. This would allow us, in principle,
to consider more general inhomogeneous cosmologies as well, not just our linearly 
perturbed FLRW case. Nevertheless, in this Letter we consider the simplest FLRW universe 
in the comoving and time orthogonal framework, which satisfies the condition that all 
domains of the scenario plotted on Fig.~\ref{Fig3}, can be described simultaneously 
within the comoving gauge condition during their evolution.

On the figure, the causal past of $D$ on $t_0$ is $J^{-}(D)$, the future development of $J^{-}(D)$ 
on $t_1$ is $F(J^{-}(D))$ (i.e.~it is the entire causally connected region to $D$ in the 
present), while $J^{-}(F(J^{-}(D)))$ is the causal past of $F(J^{-}(D))$ on $t_0$. Now,
according to definition (\ref{lambda(t)}), the ${\lambda}(t)$-parameter can be computed as
\begin{equation}\label{amax}
 {\lambda}(t)=-{\lambda}_0\,\frac{V_{J^{-}(\mathcal{D})}}{V_{J^{-}(F(J^{-}(D)))}}\,. 
\end{equation}
The most interesting feature of this result is that $\lambda(t)$ above can be approximated
by a constant with very high accuracy. Indeed, as time passes, the regions $J^{-}(D)$ and 
$J^{-}(F(J^{-}(D)))$ on $t_0$ expand with the same $\sim ct$ rate, and hence their volume 
ratio (\ref{amax}) remains approximately constant during the cosmic evolution because the effects
of inhomogeneities on $t_0$ can be neglected in the volume ratio. 
As a consequence: {\it every domain in the universe has an (approximately) constant
${\lambda}$-parameter in its relative information entropy function which can account for 
the maximal amount of information that is entangled (or mutual) between the domain and its 
causal surroundings in the model}. 

The volume ratio in (\ref{amax}) can be roughly estimated for any region below 
the size of the statistically homogeneous scale today. Considering that $t_1-t_0$ is around $13$ 
billion years since the time of decoupling, the radius of the cosmological past of 
any domain in this scale can be neglected compared to the radii $\sim c(t_1-t_0)$ 
of $J^{-}(D)$ and $\sim 2c(t_1-t_0)$ of $J^{-}(F(J^{-}(D)))$ on $t_0$. The volume 
ratio is proportional to the cubes of 
these radii so $V_{J^{-}(\mathcal{D})}/V_{J^{-}(F(J^{-}(D)))}\approx1/8$, and 
thus ${\lambda}(t)\approx-{\lambda}_0/8$. The real problem is thus how to estimate 
${\lambda}_0$ for a given domain.

\section{Time evolution} 
According to our geometric model, the Tsallis extension of the KL entropy can elegantly
measure the relative information in a cosmological domain while also describing the 
information entanglement with its causally connected surroundings. It is therefore also 
expected that similarly to the KL entropy, the time derivative of this measure can also 
provide information about the commutation relation between the time evolution and the 
volume average operation inside the domain. By differentiating \eqref{ST} with respect to 
time and after doing some algebra one can show that
\begin{equation}
\label{STdot}
-\frac{\dot S_T\{\varrho\!\mid\!\overline{\varrho}\}}{V_{D}}
=\dot{\overline{\varrho_{\lambda}}}-\overline{\dot\varrho_{\lambda}}
+\frac{\dot {\lambda}}{{\lambda}}\left[\overline{\varrho\delta_{\lambda}}
-\overline{\varrho_{\lambda}\ln\frac{\varrho}{\overline{\varrho}}}\right],
\end{equation}
where we defined the {\it ${\lambda}$-deformed density field} as 
$\varrho_{\lambda} = \varrho(1+{\lambda}\delta_{\lambda})$. 
In the limit of ${\lambda}\to 0$ it obviously recovers \eqref{dSKL},
and for the most relevant ${\lambda}\approx const.$ case, it reduces to
$\dot{\overline{\varrho_{{\lambda}}}}-\overline{\dot\varrho_{{\lambda}}}$.
Hence, the Tsallis measure can also serve as the generating functional of the 
commutation relation between the time evolution and the volume average operation 
inside the domain, however it provides this relation on $\varrho_{{\lambda}}$, which 
explicitly depends on the maximal mutual information between the domain and its 
surroundings via a simple, parametric form.

\section{Independent information}
After obtaining the mutual information between causal\-ly connected regions in the 
model, one might also be interested in how to determine the independent information 
inside a given domain. In a recent paper \cite{BV}, Bir\'o and V\'an suggested a
formalism, called the "formal logarithm" method, on how to describe 
systems with nonadditive thermodynamic properties in order to be 
compatible with the standard laws of thermodynamics. They showed
that all equilibrium compatible entropy function that follows 
Abe's nonadditive composition rule \eqref{Abe} can be mapped by the function
\begin{equation}\label{formlog}
L(S)=\frac{1}{{\lambda}}\ln[1+{\lambda}H_{\lambda}(S)],
\end{equation}
to an additive one, 
i.e.~$L(S_{\mathcal{A}\otimes\mathcal{B}})=L(S_{\mathcal{A}})+L(S_{\mathcal{B}})$,
which is also a well defined, meaningful entropy measure. The best
example of this result is the standard Tsallis entropy function \cite{Tsallis1} 
(with the same composition rule \eqref{stnadd}) whose formal logarithm
turns out to be the R\'enyi entropy formula \cite{Renyi}. Based on the 
complete analogy of our problem to the standard Tsallis--R\'enyi correspondence,
let us define the {\it R\'enyi relative entropy} function for our purposes
(for more general definitions see e.g.~\cite{MML} and references therein) 
as the formal logarithm of the Tsallis relative entropy, i.e.
\begin{equation}
S_R\{p\!\mid\!\overline{p}\}\ :=\ L(S_T\{p\!\mid\!\overline{p}\})\ 
\equiv\ \frac{1}{\lambda}\ln\left[1+{\lambda}S_T\{p\!\mid\!\overline{p}\}\right], 
\end{equation}
and give its general form as
\begin{equation}
\label{renyi}
S_R\{p\!\mid\!\overline{p}\}
=\frac{1}{q-1}\ln\left[\int_D p(x)\left(\frac{p(x)}{\overline{p}(x)}\right)^{q-1}dx\right],
\end{equation}
with $q=1+\lambda$, as before.
\eqref{renyi} clearly recovers the KL measure in the $\lambda \rightarrow 0$ limit 
and satisfies all the conditions that are required from a well defined relative 
entropy measure. In our cosmological setting it reads as
\begin{equation}
\label{SRnew}
S_R\{\varrho\!\mid\!\overline{\varrho}\}
=\frac{1}{\lambda}\ln\left[1+\lambda V_{D}\overline{\varrho\,\delta_{\lambda}}\,\right],
\end{equation}
and since it is additive for composition, the corresponding mutual information, 
$I_R(\mathcal{A},\mathcal{B})$, between two domains in our model 
is always zero. Accordingly, formula \eqref{SRnew} can measure the 
\emph{independent information} inside the domain by also taking 
into account the causal relation between the domain and its surroundings
via the $\lambda(t)$-parameter. 

\section{Summary and discussion}
In this Letter, we have investigated the problem of information entanglement between causally 
connected regions of the universe, and shown that one can estimate the effect by a parametric 
relative information entropy measure using the Tsallis formula \eqref{st}. We defined the 
parameter function of the Tsallis entropy in a geometric way 
\eqref{lambda(t)} in order to describe the causal connection between the domain and its 
surroundings. For measuring the independent information inside the domain, we proposed the 
R\'enyi relative entropy \eqref{renyi}. The present value of the $\lambda(t)$ parameter for 
every given domain may be estimated via observations, in particular, the $\lambda_0$ constant 
may be possible to obtain from large scale structure data. 

As pointed out in \cite{HBM}, the commutation rule for averaging a scalar $\psi$ in cosmology 
is $\dot{\overline\psi}-\overline{\dot\psi}=\overline{\psi\delta\theta}=\overline{\delta\psi\theta}$, 
which upon substitution in \eqref{STdot} for the most relevant $\lambda\approx const.$ case, gives 
$-{\dot S_T\{\varrho\!\mid\!\overline{\varrho}\}}/{V_{D}}=\overline{\rho_{\lambda}\delta\theta}
=\overline{\delta\rho_{\lambda}\theta}$. Therefore, the relative information entropy grows in 
time for overdense (contracting) or underdense (expanding) regions, corresponding to what is 
physically expected for large enough times in cosmology. In general, see e.g.~\cite{MT}, one 
does not expect time convexity in the growth of cosmological 
inhomogeneities. This seems to be the case for models of the present universe which include dust 
and a positive cosmological constant \cite{Sussman}. However, in those models, and whenever the 
cosmic no-hair conjecture \cite{GH} holds, one expects, for large enough times, 
that both $S_T$ and $S_R$ become increasing functions of time. Explicit examples of the evolution 
of those measures for perturbed FLRW cosmologies will be presented in \cite{CzM}.

Although in this work we derived our results in a linearly perturbed FLRW approximation, we expect 
that after appropriate generalizations, our model is also suitable to describe nonlinear effects 
as well as large, exact inhomogeneities. Beyond cosmology, our parametric approach may also 
be relevant in other, relativistic entropy problems, e.g.~black hole formation, where the nonadditive 
property of the entropy function is also a longstanding problem \cite{BCz,CzI}.

\section*{Acknowledgement}
We would like to thank to one of our referees for his extremely competent report and criticisms 
which allowed us to substantially improve this Letter from its original version. The 
research leading to these results has received funding from the European Union Seventh Framework 
Programme (FP7/2007-2013) under the grant agreement No. PCOFUND-GA-2009-246542; from CMAT through 
FEDER Funds COMPETE, and also from FCT projects Est-OE/\\MAT/UI0013/2014, SFRH/BCC/105835/2014 and 
CERN/\\FP/123609/2011. During the preparation of this Letter, V.G.Cz.~has also been supported 
by the Japan Society for the Promotion of Science L14710 grant.

\end{document}